\documentclass[aip,reprint]{revtex4-1}
\pdfoutput=1
\usepackage{color} 
\usepackage{setspace}
\usepackage{amssymb,amsmath} 
\usepackage{bbm}
\usepackage{graphicx}
\usepackage{color}
\usepackage{epstopdf}
\newcommand{\be}{\begin{equation}}
\newcommand{\ee}{\end{equation}}
\newcommand{\ben}{\begin{equation*}}
\newcommand{\een}{\end{equation*}}
\newcommand{\ket}[1]{\left| {#1}  \right \rangle}
\newcommand{\bra}[1]{\langle #1 |}

\begin{document}

\title{Non-classicality of the molecular vibrations assisting exciton energy transfer at room temperature}

\author{Edward J. O'Reilly}
\author{Alexandra Olaya-Castro}
\email{a.olaya@ucl.ac.uk}
\affiliation{Department of Physics and Astronomy, University College London, Gower Street, London, WC1E 6BT United Kingdom.}
\begin{abstract}
Advancing the debate on quantum effects in light-initiated reactions in biology requires clear identification of non-classical features that these processes can exhibit and utilise. Here we show that in prototype dimers present in a variety of photosynthetic antennae, efficient vibration-assisted energy transfer in the subpicosecond timescale and at room temperature can manifest and benefit from non-classical fluctuations of collective pigment motions. Non-classicality of initially thermalised vibrations is induced via coherent exciton-vibration interactions and is unambiguously indicated by negativities in the phase-space quasi-probability distribution of the effective collective mode coupled to the electronic dynamics. These quantum effects can be prompted upon incoherent input of excitation. Our results therefore suggest that investigation of the non-classical properties of vibrational motions assisting excitation and charge transport, photoreception and chemical sensing processes could be a touchstone for revealing a role for non-trivial quantum phenomena in biology. 
\end{abstract}
\maketitle
\subsection*{Introduction}
The experimental demonstration of oscillatory electronic dynamics in light-harvesting complexes \cite{Engel07, Panitchayangkoon07,Calhoun09,Collini10, Harel12} has triggered wide-spread interest in uncovering quantum phenomena that may impact the function of the molecular components of living organisms. In general, however, oscillatory patterns in dynamics is not sufficient argument to rule out  classical descriptions of the same behaviour. Indeed, recent works have discussed how classical coherence models can predict electronic coherence beating \cite{Miller12, Briggs11}.  Therefore, an important challenge for the growing field of quantum effects in biomolecules is to clearly identify which quantum features with no classical counterpart may manifest in these systems and how they may influence the process of interest. 

The question of non-classicality of the dynamics of electronic excitations in light-harvesting systems has been addressed by investigating Leggett-Garg inequalities \cite{Wilde10}. This work concludes that, under Markovian evolution, temporal correlations of individual pigments observables should violate classical bounds, and in consequence certain classical theories are unsuitable to describe electronic dynamics. Other works have investigated the quantumness of the electronic degrees of freedom \cite{Olaya08,Sarovar10,Caruso10,Fassioli10, Whaley11}. Despite these efforts, it is still far from understood which non-classical phenomena are directly correlated  with efficient energy distribution in a prototype light-harvesting system.

What is clear is that exciton energy transport depends not only on the topology of electronic couplings among pigments but is critically determined by exciton-phonon interactions: molecular motions \cite{Renger01} and environmental fluctuations \cite{Renger01, Plenio08, Rebentrost09} drive efficient transport processes in light-harvesting antennae. In fact, it is well known that exciton-phonon interactions in these complexes have a rich structure as a function of energy and generally include coupling to both continuous and discrete modes  associated to low-energy solvated protein fluctuations and underdamped intramolecular vibrations, respectively \cite{Renger01}. Moreover, evidence is mounting that the interaction between excitons and underdamped vibrations whose energies commensurate exciton splittings may be at the heart of the coherence beating probed in two-dimensional photon echo spectroscopy \cite{Richards12,Turner12,Kolli12,Chin12a,Chin12b, Christensson12, Tiwari12}.  Although some insights into the importance of such resonances can be gained from  F\"orster theory \cite{Forster59}, the wider implications for optimal spatio-temporal distribution of energy \cite{Perdomo10, Kolli12,Rey13}, for modulation of exciton coherences \cite{Kolli12, Chin12a, Chin12b, Christensson12} and for collective pigment motion dynamics \cite{Tiwari12} have just recently started to be clarified. 

The current state of the debate then suggests that a conceptual advance in understanding non-trivial quantum phenomena assisting electronic transport could emerge precisely from investigating non-classical features of the molecular motions and phonon environments that play such a key role. Techniques able to manipulate vibrational states \cite{Biggs09,Biggs12} and probe their quantum properties \cite{Waldermann08, Lee11} may indeed provide the experimental platform to address this issue. Here, we investigate this question in a prototype dimer ubiquitous in light-harvesting antennae of cyanobacteria \cite{Womick11}, cryptophyte algae \cite{Doust04,Novoderezhkin10} and higher plants \cite{Liu04, Barros09, Novoderezhkin04b} and show that commensurate energies of exciton splitting and underdamped  high-energy vibrations allows exciton-vibration dynamics to induce and harness non-classical fluctuations of collective pigment motions for efficient energy transfer. Negative values of the Mandel $Q$-parameter \cite{Mandel79} indicating sub-Poissonian phonon occupation fluctuations and, correspondingly, negative regions in a regularised quasi-probability $P$ distribution in phase space \cite{Kiesel10, Kiesel11}, unambiguously preclude any classical description of such fluctuations or its correlations with transport.  Our results show a potential functional relevance of non-classicality of molecular fluctuations for exciton transport and therefore provide a framework to investigate similar  non-trivial quantum phenomena in the large variety of biomolecular transport \cite{Novoderezhkin04a, Hay12}, photoreception \cite{Dasgupta09} and chemical sensing processes \cite{Turin96, Brookes07,Franco11} that are known (or hypothesised) to be assisted by unequilibrated vibrational motion.

\subsection*{Results}
{\bf Characterising non-classicality. } The field of quantum optics has developed a solid framework to quantify the quantum properties of bosonic fields \cite{Davidovich96}. It therefore provides excellent conceptual and quantitative tools to investigate non-classicality of the harmonic vibrational degrees of freedom of interest in this work. From the perspective of quantum optics, quantum behaviour with no classical counterpart i.e. on-classicality, arises if the state of the system of interest cannot be expressed as a statistical mixture of coherent states defining a valid probability measure \cite{Scully97}. This then leads to non-positive values of a phase-space quasiprobability distribution such as the Glauber-Sudarshan $P (\alpha)$-function\cite{Scully97}.   
\begin{equation}
P(\alpha)  = \frac{1}{\pi^2}\int d^2\xi \, e^{\alpha \xi^*-\alpha^*\xi} \chi(\xi) \, ,
\end{equation}
where $\chi(\xi)$ is the characteristic function of the bosonic quantum state.  However, highly singular behaviour of $P(\alpha)$ can make its characterisation challenging both theoretically and experimentally. To overcome this, verification of the non-classicality of a quantum state can be done by constructing a regularised distribution  $P_w(\alpha)$ as the Fourier transform of a filtered quantum characteristic function $\chi_w(\xi)$  \cite{Kiesel10, Kiesel11} as explained in the methods section. Negativities in this regularised distribution are necessary and sufficient condition of quantum behaviour with no classical analogue \cite{Kiesel10} and offer a significant advantage over other distributions, such as the Wigner distribution, which can be positive for quantum states that are truly non-classical.  

As an alternative to phase-space distributions, signatures of non-classicality can be observed in the fluctuations of the bosonic field. For a single-mode, negative values of the Mandel's $Q-$parameter \cite{Mandel79} can be a signature of non-classical behaviour.  It characterises  the departure of the occupation number distribution $P(n)$ from Poissonian statistics through the inequality
\begin{eqnarray}
Q=\frac{\langle \hat n^2 \rangle- \langle \hat n \rangle^2}{\langle \hat n \rangle} - 1 <0
\end{eqnarray}
where $\langle \hat n \rangle$ and $\langle \hat n^2 \rangle$ denote the first and second moments of the bosonic number operator $\hat n$ respectively. Vanishing $Q$ indicates Poissonian number statistics where the mean of $\hat n$ equals its variance as it is characteristic of classical wave-like behaviour i.e. a coherent state of light.  For a chaotic thermal state one finds that $Q=\langle \hat n \rangle >0$ indicating that particles are `bunched'.  A Fock state is characterised by $Q< 0$ indicating  that particle occupation is restricted to a particular level.  Inequalities involving occupation probabilities of nearest number states can similarly witness  non-classical occupation fluctuations \cite{Klyshko96}.

In what follows we use the above framework to investigate the non-classical behaviour of the vibrational motions that drive excitation dynamics in prototype dimers present in a variety of antennae proteins of photosynthetic systems. 
\begin{figure}[tbp]
\includegraphics{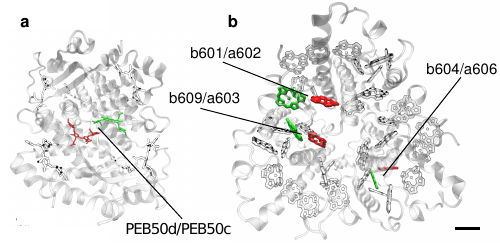}
\caption{\textbf{Prototype dimers}.  (\textbf{a}-\textbf{b}) Cryptophyte antennae phycoerythrin 545 (PE545) and light harvesting complex II (LHCII) present in higher plants have pairs of pigments whose electronic and vibrational parameters fall in the regime of our vibration-assisted transport model. (\textbf{a}) Representation of the pigments and protein environment of  a PE545 complex of \textit{Rhodomonas} CS24 (Protein Data Bank ID code \texttt{1XG0} (ref. \citenum{Doust04})). The central phycoeritrobilin (PEB) dimer pigments $\textrm{PEB}_{50c}$ and $\textrm{PEB}_{50d}$ are highlighted red and green respectively. For this $\textrm{PEB}_{50}$ dimer there is an uncertainty in the value of the energy gap  \cite {Doust04, Novoderezhkin10}. We take parameters from \cite{Doust04, Novoderezhkin10} such that $\Delta \varepsilon= 1042 ~\textrm{cm}^{-1}$ and $V=92~{\rm cm}^{-1}$ so $\Delta E=1058.2~\textrm{cm}^{-1}$ being quasiresonant with an intramolecular mode of frequency $\omega_\textrm{vib}=1111~\textrm{cm}^{-1}$. The strength of linear coupling to this mode is $g=\omega_\textrm{vib}\sqrt{0.0578}=267.1~\textrm{cm}^{-1}$.   (\textbf{b}) Representation of the LHCII antennae of \textit{Spinacia oleracea} (Protein Data Bank ID code \texttt{1RWT} (ref. \citenum{Liu04}). Several pairs of close $\textrm{Chl}_b$-$\textrm{Chl}_a$ (red-green) chlorophylls satisfy the conditions of our model. In particular, we consider the $\textrm{Chl}_{b601}$-$\textrm{Chl}_{a602}$ pair for which $\Delta \varepsilon=661~\textrm{cm}^{-1}$ and $V=-47.1~\textrm{cm}^{-1}$, resulting in $\Delta E =667.7~\textrm{cm}^{-1}$ \cite{SchlauCohen09}.  An intramolecular vibrational mode of frequency $\omega_\textrm{vib}=742.0~\textrm{cm}^{-1}$ is close to this energy gap and each chromophore couples to this mode with strength $g=\omega_\textrm{vib}\sqrt{0.03942}=147.3~\textrm{cm}^{-1}$ as obtained from ref.~\citenum{Novoderezhkin04b}. Scale bar: 1~nm.
}
\label{fig:prototype}
\end{figure}
\begin{figure}[htbp]
\includegraphics{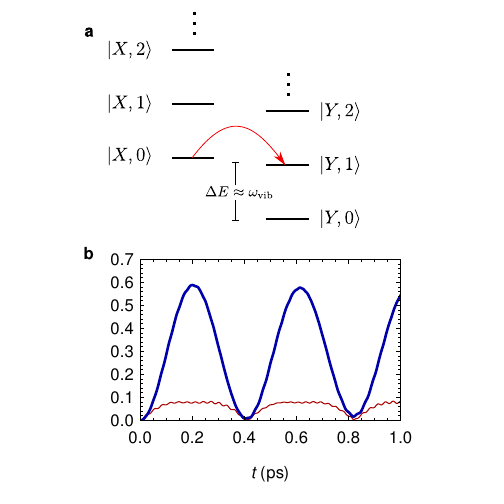}
\caption{\textbf{Exciton-collective mode states and free exciton dynamics}. (\textbf{a}) The energy levels of the exciton-collective mode states used to describe energy transfer $\ket{X,n}$. The red arrow denotes population transfer from $\ket{X,0}$ to $\ket{Y,1}$. (\textbf{b}) Quantum coherent dynamics of the $\textrm{PEB}_{50}$ dimer in PE545 illustrating population of the lowest exciton  $\rho_{YY}(t)$ (thick blue curve) and the inter-exciton coherence in the ground-state of the collective mode $|\rho_{X0-Y0}(t)|$ (thin red curve).}
\label{fig:freedynel}
\end{figure}

\begin{figure*}[htbp]
\centering
\includegraphics{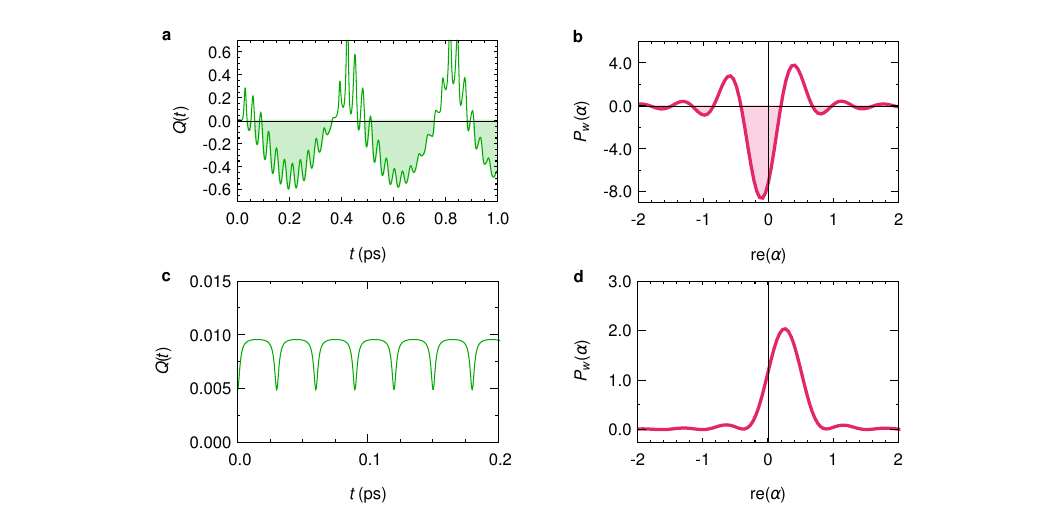}
\caption{\textbf{Non-classicality of collective and local vibrational modes }.  (\textbf{a}) Mandel $Q$-parameter of the relative displacement mode of the considered $\textrm{PEB}_{50}$ dimer from  PE545 when a biological electronic coupling is considered. Shaded regions denote times of non-classicality. (\textbf{b}) The associated regularised quasiprobability distribution $P_w(\alpha)$ at $t=0.2~$ps. Shaded regions denote areas of negative probability.
(\textbf{c}) Mandel $Q$-parameter of the intramolecular high-energy vibration when dipole coupling is zero and (\textbf{d}) the associated regularised quasiprobability distributions $P_w(\alpha)$ at $t=0.2~$ps.
}
 \label{fig:freedynmode}
\end{figure*}

{\bf Prototype dimers and collective motion. } 
We consider a prototype dimer where each chromophore has an excited electronic state of energy $\varepsilon _i$ strongly coupled to a quantized vibrational mode of frequency $\omega_\textrm{vib}$ much larger than the thermal energy scale $K_\textrm{B}T$ and described by the bare Hamiltonians
$ H_{\textrm{el}} = ~\sum_{i=1,2}\varepsilon_i\sigma^+_i\sigma^-_i$
and 
$H_{\textrm{vib}}= ~\omega_{\textrm{vib}}(b_1^\dag b_1+ ~b_2^\dag b_2)$
respectively.  Inter-chromophore coupling is generated by dipole-dipole interactions of the form 
$ H_{\textrm{d-d}} = V(\sigma^+_1\sigma^-_2+\sigma^+_2\sigma^-_1)$. 
The electronic excited states interact with their local vibrational environments with strength $g$, linearly displacing the corresponding mode coordinate,
$ H_{\textrm{el-vib}} = g \sum_{i=1,2}\sigma^+_i\sigma^-_i(b^\dag_i+ b_i)$.
In the above $b^\dag_i$($b_i$) creates (annihilates) a phonon of the vibrational mode of chromophore $i$ while $\sigma^\pm_i$ creates or destroys an electronic excitation at site $i$. The eigenstates $|X\rangle$ and $|Y\rangle$ of 
$H_{\textrm{el}}+H_{\textrm{d-d}} $ denote exciton states with energy splitting given by 
$\Delta E =\sqrt{(\Delta \varepsilon)^2+4V^2}$ and $ \Delta \varepsilon=\varepsilon_1-\varepsilon_2$.
Transformation into collective mode coordinates shows that centre of mass mode $b^{(\dag)}_\textrm{cm}=(b^{(\dag)}_1+b^{(\dag)}_2)/\sqrt{2}$ decouples from the electronic degrees  of freedom and that only the 
mode corresponding to the relative displacement mode with  bosonic operators
\begin{equation}
b_\textrm{rd}^{(\dag)}=(b^{(\dag)}_1- b^{(\dag)}_2)/\sqrt{2},
\end{equation}
couples to the excitonic system.  It is the non-classical properties of  this collective mode which we investigate. In collective coordinates, the effective exciton-vibration Hamiltonian then reads 
\begin{equation}
 H_{\textrm{ex-vib}} = \frac{\Delta \varepsilon}{2} \sigma_z +  V \sigma_x - \frac{g}{\sqrt 2} \sigma_z (b_{\textrm{rd}}^{\dag}+b_\textrm{rd}) +\omega_{\textrm{vib}} b^{\dag}_\textrm{rd}b_{\textrm{rd}}, 
\end{equation}
with $\sigma_z=\sigma^+_2\sigma^-_2-\sigma^+_1\sigma^-_1$ and $\sigma_x=(\sigma^+_1\sigma^-_2+\sigma^+_2\sigma^-_1)$.  Tiwary and co-authors \cite{Tiwari12} have recently pointed out that two-dimensional spectroscopy can probe the involvement of these anti-correlated, relative displacement motions in electronic dynamics. From now on and for simplicity we denominate this relative displacement mode as the collective mode.  

We are interested in dimers that satisfy $\Delta E \sim\omega_{\textrm{vib}}>g>V$ where the effects of underdamped high-energy vibrational motions are expected to be most important \cite{Womick11, Kolli12, Novoderezhkin04b}. Several natural light-harvesting antennae include pairs of chromophores that clearly fall in this regime. Two important examples of such dimers are illustrated in Fig. \ref{fig:prototype}a-b  and correspond to the central $\textrm{PEB}_{50c}$-$\textrm{PEB}_{50d}$ dimer in the cryptophyte antennae PE545 \cite{Novoderezhkin10} and a $\textrm{Chl}_{b601}$-$\textrm{Chl}_{a602}$ pair in the LHCII complex of higher plants \cite{Novoderezhkin04b}; both corresponding to systems which have exhibited coherence beating in two-dimensional spectroscopy  \cite{Collini10, Wong12, Calhoun09}.  Importantly, in each case, the dimer considered contributes to an important energy transfer pathway towards exit sites \cite {Novoderezhkin10, Calhoun09} suggesting that the the phenomena we discuss will have an effect in the performance of the whole complex. Moreover, synthetic versions of such prototype  dimers could be available  \cite{Hayes13}. Most remarkably, LHCII is likely the most abundant light-harvesting complex on Earth \cite{Barros09}, while cryptophyte antennae like PE545 are ecologically important as they support photosynthesis under extreme low-light conditions \cite{Samsonoff01,Scholes12b}. From this perspective,  the dimers of interest are exceptionally relevant biomolecular prototypes. Spectroscopy studies indicate that these dimers are subject to a structured exciton-phonon interaction as considered in our model.  For the $\textrm{PEB}_{50}$ dimer, the intramolecular mode of interest has frequency around 1111 cm$^{-1}$ \cite{Novoderezhkin10} which compares with the frequency of the breathing mode of the tetrapyrrole\cite{Singh08} (Carles Curutchet, personal communication). In the case of $\textrm{Chl}_{b-a}$ pair it has been shown that a mode around 750 cm$^{-1}$ is coupled to the electronic dynamics \cite{Novoderezhkin04b} and this energy is close to the frequency of in-plane deformations of the pyrrole \cite{Zhou97}. Furthermore, vibrational dephasing in chromophores \cite{Vohringer95} and in other systems such as photoreceptors \cite{Kukura07} is known to be of the order of picoseconds.  Some aspects of the influence of non-equilibrium vibrational motion in these specific dimers have been considered before \cite{Kuhn96, Kolli12}, but none of these works have addressed the question of interest: can vibration-assisted transport exploit quantum  phenomena that has no classical analogue? 

{\bf Non-classicality via coherent exciton-vibration dynamics. }
We first consider the quantum coherent dynamics associated to $H_{\textrm{ex-vib}}$ to illustrate how non-classical behaviour of the collective motion emerges out of an initial thermal phonon distribution and an excitonic state with no initial superpositions: $\rho(t_0) = \ket{X}\bra{X}\otimes\varrho^\textrm{th}_\textrm{vib}$ which written in the basis of exciton-vibration states of the form $|X,n\rangle$ (see Fig. \ref{fig:freedynel}a) becomes $\rho(t_0) = \sum_nP_\textrm{th}(n) \ket{X,n}\bra{X,n}$. Here $n$ denotes the phonon occupation number of the relative displacement mode coupled to exciton dynamics (see Eq. 4) while $P_\textrm{th}(n)$ denotes the thermal occupation of such level. The observables of interest are  the population of the lowest excitonic state  
$\rho_{YY}(t)= \sum_n \langle Y,n| \rho(t)|Y,n \rangle $,
the absolute value of the coherence $\rho_{X0-Y0}(t)=\langle X,0| \rho(t)|Y,0 \rangle$ which denotes the inter-exciton coherence in the ground-state of the collective vibrational mode, and the non-classicality given by negative values $Q(t)$ and corresponding negativities in the regularised quasi-probability distribution $P_w(\alpha)$.  Hamiltonian evolution generates coherent transitions from states $|X,n\rangle$ to $|Y,n+1\rangle$ (see Fig. \ref{fig:freedynel}a) with a rate $f$ that depends on the exciton delocalization ($|V|/\Delta \varepsilon$), the coupling to the mode $g$, and the phonon occupation $n$ i.e. $f \simeq g (2|V|/\Delta \varepsilon)\sqrt{(n+1)/2}$. Since $\omega_\textrm{vib} \gg K_\textrm{B}T$ the ground state of the collective mode is largely populated, such that the Hamiltonian evolution of the initial state is dominated by the evolution of  the state $|X,0\rangle$.  This implies that the energetically close exciton-vibration state $|Y,1\rangle$ becomes coherently populated at a rate $f\simeq g (2|V|/\Delta \epsilon)$, leading to the oscillatory pattern observed in the probability of occupation $\rho_{YY}(t)$ as illustrated in Fig. \ref{fig:freedynel}b. The low-frequency oscillations of the dynamics of $\rho_{YY}(t)$ cannot be assigned to the exciton or the vibrational degrees of freedom alone as expected from quantum-coherent evolution of the exciton-plus-effective mode system. For instance, if the mode occupation is restricted to at most $n=1$, the period of the amplitude of  $\rho_{YY}(t)$ is given approximately by  the inverse of 
\be
\frac{1}{2}\left(\sqrt{(\Delta E-\alpha)^2+\frac{2g^24V^2}{\alpha \Delta E }}-(\Delta E+\alpha)\right)~,
\ee
with $\alpha^2=2g^2+\omega_\textrm{vib}^2$ and $(2g^24V^2/\alpha^2\Delta E^2)\ll 1$. Coherent exciton population transfer is accompanied by beating of the inter-exciton coherence  $|\rho_{X0-Y0}(t)|$ with the main amplitude modulated by the same low-frequency oscillations of $\rho_{YY}(t)$ and a superimposed fast oscillatory component of frequency close to $\omega_{\textrm{vib}}$ (see Fig. \ref{fig:freedynel}b). This  fast driving component arises from local oscillatory displacements: when $V\simeq 0$ the time evolution of each local mode is determined by the displacement operator with amplitude  $\alpha(t)={2g}(1-\exp(-i\omega_\textrm{vib} t))/\omega_\textrm{vib}$ \cite{Palma96}. As the state $|Y,1\rangle$ is coherently populated, the collective quantised mode is driven out of equilibrium towards a non-classical state in which selective occupation of the first vibrational level takes place thereby modulating occupation of higher levels. This manifests itself in sub-Poissonian phonon statistics as indicated by negative values of $Q(t)$ shown in Fig. \ref{fig:freedynmode}a. Similar phenomena have been described in the context of electron transport in a nanoelectromechanical system \cite{Merlo08,Cavaliere08}. Moreover, Fig. \ref{fig:freedynmode}b shows that at times when $Q(t)$ is negative i.e. $t=0.2$ ps the regularized quasiprobability distribution $P_w(\alpha)$ at this time exhibits negativities thereby ruling out any classical description of the same phenomena. Interestingly, the non-classical properties of the collective vibrational motion resemble non-classicality of bosonic thermal states (completely incoherent states) that are excited by a single quanta \cite{Zavatta07, Kiesel11}. 
Importantly, such non-classical behaviour of the vibrational motion arises only when the electronic interaction between pigments is finite. For comparison, Fig. \ref{fig:freedynmode}c shows that if $V=0$, an electronic excitation drives the local underdamped vibration towards a thermal displaced state with super-Poissonian statistics ($Q(t)>0$) which has an associated positive probability distribution in phase space as illustrated in Fig. \ref{fig:freedynmode}d.  In short, non-classicality of the collective mode quasiresonant with the excitonic transition arises through the transient formation of exciton-vibration states. 

\begin{figure*}[htbp]
\centering
\includegraphics{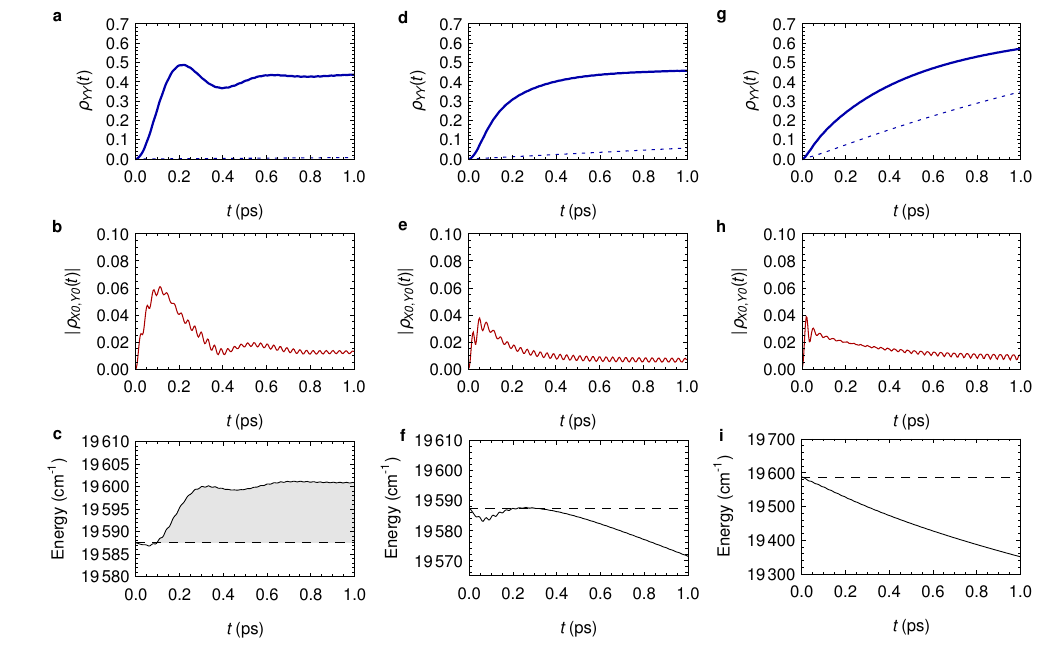}
\caption{\textbf{Energy and coherence evolution under thermal relaxation}. The dynamics of $\rho_{YY}(t)$ with (thick blue curve) and without (dashed curve) coupling to vibration (top row), $|\rho_{X0,Y0}(t)|$ (middle row) and energy of the exciton vibration system $E(t)=\textrm{Tr}\{H_\textrm{ex-vib} \rho(t)\}$ (bottom row) for the exciton-vibration parameters of the $\textrm{PEB}_{50}$ dimer and three interaction strengths to the low-energy thermal bath: (\textbf{a}-\textbf{c}): $\lambda = 6~\textrm{cm}^{-1}$, (\textbf{d}-\textbf{f}) $\lambda = 35~\textrm{cm}^{-1}$ and  (\textbf{g}-\textbf{i}): $\lambda = 110~\textrm{cm}^{-1}$. Initial energy $E(0)$, displayed as dashed line and times where $E(t)>E(0)$ are shaded.}
\label{fig:dynamics}
\end{figure*}
 {\bf Energy and coherence dynamics under thermal relaxation. } 
 We now investigate the dynamics of the exciton-vibration dimer when each local electronic excitation interacts additionally with a  low-energy thermal bath described by a continuous distribution of harmonic modes. The strength of this interaction is described by a Drude spectral density with associated  reorganisation energy $\lambda$ and cut-off frequency $\Omega_\textrm{c} < K_\textrm{B}T$ as described in the methods section. We consider the exciton and vibration parameters to the $\textrm{PEB}_{50}$ dimer and investigate the trends as functions of the reorganisation energy. As expected, the interplay between vibration-activated dynamics and thermal fluctuations leads to two distinct regimes of energy transport as a function of $\lambda$. For our consideration of weak electronic coupling, the coherent transport regime is determined approximately by  $\sqrt{\lambda \Omega_\textrm{c}} \leq 2g|V|/\Delta \epsilon$. Population of the low-lying exciton state is dominated by coherent transitions between exciton-collective mode states and the rapid, non-exponential growth of  $\rho_{YY}(t)$ in this regime can be traced back to coherent evolution from $|X,0\rangle$ to $|Y,1\rangle$. At longer time scales thermal fluctuations induce incoherent transitions from $|X,0\rangle$ to $|Y,0\rangle$ with a rate  proportional to $\sqrt{\lambda \Omega_\textrm{c}}$, thereby stabilising population of $\rho_{YY}(t)$ to a particular value as can be seen in Fig. \ref{fig:dynamics}d.  This behaviour is illustrative of what is expected in the dimer $\textrm{Chl}_{b-a}$ for which $\lambda=37~$cm$^{-1}$ as obtained from ref. \citenum{Novoderezhkin04b}. To confirm this we have computed the exciton-vibration dynamics with parameters of the  $\textrm{Chl}_{b-a}$ dimer, the results of which are shown in Supplementary Fig. S1.
In contrast,  for $\sqrt{\lambda \Omega_\textrm{c}}>2g |V|/\Delta \epsilon$ population  transfer to $\rho_{YY}(t)$ is incoherent.  For the $\textrm{PEB}_{50}$ dimer $\lambda \sim 110 {\rm cm}^{-1}$ which place this dimer in this incoherent regime where $\rho_{YY}(t)$ has a slow but continuous exponential rise reflecting the fact that thermal fluctuations inducing transitions from $|X,n\rangle$ to $|Y,n\rangle $ now have a large contribution to exciton transport.  However, even in this regime, transfer to $\rho_{YY}(t)$ is always more efficient with the quasi-resonant mode than in the situations where only  thermal-bath induced transitions are considered (see dashed lines in Fig. \ref{fig:dynamics}a, d and g). The underlying reason is that before vibrational relaxation takes place  (around $t=1~\textrm{ps}$), the system is transiently evolving towards a thermal configuration of exciton-collective mode states. Hence,  in both coherent and incoherent population transfer regimes transfer to the lowest exciton state involves a transient, selective population of  first vibrational level of the collective mode.  The transition from coherent to incoherent exciton population dynamics is then marked by the onset of energy dissipation of the exciton-vibration system as shown in Fig. \ref{fig:dynamics}c, f and i where $E(t)=\textrm{Tr}\{H_\textrm{ex-vib}\rho(t)\}$ has been depicted for different values of $\lambda$. While exciton population growth is non-exponential, energy dissipation into the thermal bath is transiently prevented  as indicated by periods of positive slope of $E(t)$ as happens in Fig. \ref{fig:dynamics}c and \ref{fig:dynamics}f.  Quantification of the energy that is transiently ``extracted" from the low-energy thermal bath can provide an interesting physical interpretation of the advantages of non-exponential exciton transfer in the framework of non-equilibrium thermodynamics \cite{Jianming10}.
\begin{figure*}[htbp]
\centering
\includegraphics{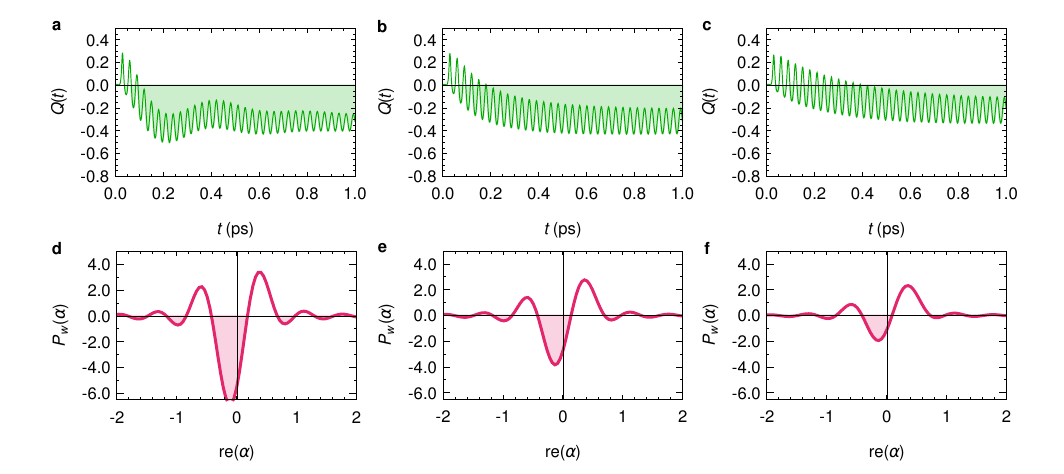}
\caption{\textbf{Non-classicality of collective motions under thermal relaxation}. (\textbf{a-c}) Dynamics of the Mandel $Q$-parameter for $\lambda = 6,35,110~\textrm{cm}^{-1}$ respectively. Shaded regions denote times of non-classicality. (\textbf{d-f}) Regularised distribution $P_w(\alpha)$ at $t=0.2$ ps for each corresponding value of $\lambda$. Shaded regions denote areas of negative probability.}
\label{fig:mandel}
\end{figure*}
For completeness, we present in Fig. \ref{fig:dynamics}b, e  and h how the beating patterns of the coherence  $\rho_{X0-Y0}(t)$ reveal the structured nature of the exciton-phonon interaction and witnesses whether there is coherent exciton-vibration evolution as it has been pointed out by recent studies \cite{Christensson12, Chin12b}. The frequency components of such oscillatory exciton coherences vary depending of the coupling to the thermal bath. In the coherent regime, as $\rho_{X0-Y0}(t)$  follows exciton populations, the main amplitude is modulated by the same relevant energy difference between exciton-vibration states (see Fig.  \ref{fig:dynamics}b and e). This behaviour is relevant for the parameter regime of the  $\textrm{Chl}_{b-a}$ dimer (see Supplementary Fig. S1). In contrast, for the  $\textrm{PEB}_{50}$ dimer, the short-time oscillations of $\rho_{X0-Y0}(t)$ (between $t=0$ and $t=0.1$ps) arise from purely electronic correlations due to bath-induced renormalization of the electronic Hamiltonian \cite{Silbey84}.  This exciton coherence retains the superimposed driving at a frequency $\omega_\textrm{vib}$  and is accompanied by non-classicality as it will be described shortly, indicating that vibrational motion is still out of thermal equilibrium. The dynamical features presented in Fig. \ref{fig:dynamics}g and h agree with previous findings based on a perturbative approach \cite{Kolli12}  and with the time-scales of the exciton coherence beating reported for cryptophyte algae \cite{Collini10,Turner12}. 

{\bf Non-classicality under thermal relaxation. }  
Interaction with the thermal environment would eventually lead to the emergence of classicality in longer time scales. However, in the picosecond time scale of interest, the collective mode exhibits periods of non-classicality across a wide  range of thermal bath couplings $\lambda$ as indicated by sub-Possonian fluctuations with $Q(t)<0$ in Fig. \ref{fig:mandel}a-c and the corresponding negativities in the distributions $P_w(\alpha)$ shown in Fig. \ref{fig:mandel}d-f.  This survival of non-classicality is concomitant with a slow decay of the  exciton-vibration coherence $\rho_{X0,Y1}(t)$ (not shown). Non-classical behaviour of collective fluctuations are then expected  for the parameters of both the $\textrm{PEB}_{50}$ dimer for which $\lambda=110 ~\textrm{cm}^{-1}$ and the $\textrm{Chl}_{b-a}$ dimer for which $\lambda=37 ~\textrm{cm}^{-1}$. The non-classical  fluctuations predicted by $Q(t)$  also agree with those witnessed by a parameter quantifying correlations between nearest-neighbours occupations  \cite{Klyshko96} which we present Supplementary Fig. S2.  As expected, the maximum non-classicality indicated by the most negative value of $Q(t)$ decreases for larger reorganisation energies. Nonetheless,  the time average of these non-classicality is not a monotonic function of $\lambda$.  For moderate  values of $\lambda$, the collective  mode spends longer periods in states with non-classical fluctuations i.e. periods for which $Q(t)<0$ as seen in Fig. \ref{fig:mandel}b thereby stabilising non-classicality at a particular level. This sub-picosecond stabilisation of non-classicality is expected in the regime of the $\textrm{Chl}_{b-a}$ dimers as illustrated in Supplementary Fig. S1.

\begin{figure}[tbp]
\centering
\includegraphics{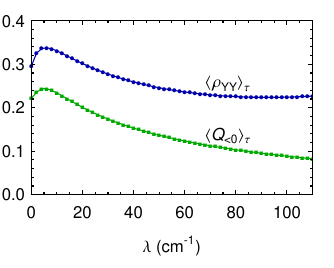}
\caption{\textbf{Correlations between non-classicality and exciton populations}. 
Time integrated averages of exciton population $\rho_{YY}(t)$ and non-classicality as quantified by $Q(t)\Theta[-Q(t)]$ (blue and green respectively) as functions of coupling to the thermal background by fixing environment cut-off frequency $\Omega_\textrm{c} = 100~\textrm{cm}^{-1}$ and varying reorganization energy $\lambda$ with exciton-vibration parameters corresponding to the $\textrm{PEB}_{50}$ dimer.
}
\label{fig:averages}
\end{figure}

{\bf Functional role of non-classicality. }
Non-classical fluctuations of collective motions correlate to exciton population transfer. In order to demonstrate this, we investigate quantitative relations between non-classicality and exciton energy transport by considering relevant integrated averages in the time scale of the Hamiltonian evolution of the exciton-vibration system denoted by $\tau$. For the parameters of the $\textrm{PEB}_{50}$, this time scale is about half a picosecond and is comparable to the time scale in which excitation energy would be distributed away to other chromophores or to a trapping state.  The time integrated averages over $\tau$ are defined as: 
$
\langle F [\rho(t)]\rangle_{\tau}=\frac{1}{\tau}\int_0^\tau\textrm{d}t~ F[\rho(t)]~,
$
where $F[\rho(t)]$ corresponds to the exciton population $\rho_{YY}(t)$ and the non-classicality of the underdamped collective mode through periods of sub-Poissonian statistics $Q(t)\Theta[-Q(t)]$  as functions of the coupling to the bath $\lambda$. As shown in Fig. \ref{fig:averages} the average exciton population and non-classicality follow a similar non-monotonic trend as a function of the coupling to th e thermal bath, indicating a direct quantitative relation between efficient energy transfer in the time scale $\tau$ and the degree of non-classicality. The appearance of a maximal point in the average non-classicality as a function of the system-bath coupling indicates that the average quantum response of the collective anti-correlated motion to the impulsive electronic excitation, is optimal for a small amount of thermal noise. 

\begin{figure}[]
\centering
\includegraphics{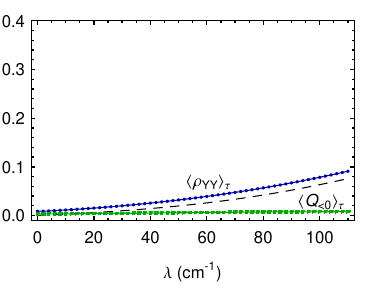}
\caption{\textbf{Effects of off-resonance vibrations.} Time integrated averages as in Fig. \ref{fig:averages} with electronic parameters of the $\textrm{PEB}_{50}$ dimer but now considering and intramolecular vibration which $\omega_\textrm{vib}=1520~\textrm{cm}^{-1}$ \textit{i.e.} significantly detuned from the energy gap $\Delta E=1058.2~\textrm{cm}^{-1}$ and similar coupling strength $g=\omega_\textrm{vib}\sqrt{0.0265}\approx 247~\textrm{cm}^{-1}$. Parameters obtained from ref. \citenum{Novoderezhkin10}. The dashed curve shows time average of $\rho_{YY}(t)$ when no intramolecular vibration is considered and transport results only due to the thermal background.
}
\label{fig:offresonant}
\end{figure}

It is worth emphasising that the above functional role of non-classicality holds for vibration-assisted transport where the high-energy intramolecular modes considered are quasi-resonant with the excitonic energy splitting. When vibrational motions are significantly detuned with the bare exciton transition $\Delta E$, transport is dominated by the thermal background and no selective population of the state $|Y,1\rangle$ takes place; hence periods of sub-Poissonian fluctuations vanish. To illustrate the difference with the off-resonance case Fig. \ref{fig:offresonant} shows the same time integrated averages as in Fig. \ref{fig:averages} with the electronic parameters of the $\textrm{PEB}_{50}$ dimer but now  considering an intramolecular vibration of frequency $\omega_{\rm vib}=1520 {\rm cm}^{-1}$ significantly detuned from $\Delta E$. In this case thermally activated transport (see dashed line in Fig. \ref{fig:offresonant}) and vibration-assisted transport  (solid line in Fig. \ref{fig:offresonant}) are practically indistinguishable and the average $Q(t)$ is positive with a value near zero as expected for a thermal distribution of a high-energy harmonic mode. 

\begin{figure}[htbp]
\centering
\includegraphics{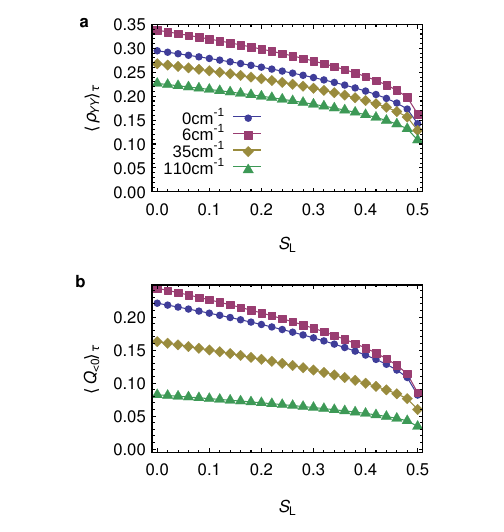}
\caption{\textbf{Non-classicality under incoherent exciton input.} 
The time integrated averages of exciton population $\rho_{YY}(t)-\rho_{YY}(0$) and non-classicality as quantified by $Q(t)\Theta[-Q(t)]$ as functions of the linear entropy $S_\textrm{L}=2r(r-1)$ for $0<r<1/2$. Curves are for different system-bath couplings. 
}
\label{fig:incoherent}
\end{figure}

The degree of purity of the initial exciton state is also important to enabling and harnessing non-classical fluctuations of the collective mode. One therefore should expect that statistical mixtures of excitons with finite purity can still trigger such non-classical response. To illustrate this we now consider mixed initial states of the form $\rho(t_0) = \rho_{\textrm{ex}}\otimes\varrho^\textrm{th}_\textrm{vib}$ where $\rho_{\textrm{ex}}=r\ket{X}\bra{X}-(1-r)\ket{Y}\bra{Y}$ with $1/2\leq r\leq 1$. The associated linear entropy quantifying the mixedness of the initial exciton states is given by $S_\textrm{L}=2r(r-1)$. The time-averaged non-classicality (Fig. \ref{fig:incoherent}a) and average population transfer (Fig. \ref{fig:incoherent}b) follow similar decreasing, yet non-zero, monotonic trends for mixed states. These results suggest that non-classical vibrational motion can be prompted and exploited even under incoherent conditions creating statistical mixture of excitons \cite{Fassioli12}. The trends presented in Figures \ref{fig:averages} and  \ref{fig:incoherent} constitute theoretical evidence of direct quantitative correlations between non-classicality and exciton population transport in a relevant sub-picosecond time scale and therefore  illustrate a functional role for quantum phenomena with no classical counterpart in a prototype light-harvesting system.  Our results remain valid when the picosecond dephasing rate of the vibrational motion is included in the dynamics as can be seen in Supplementary Fig. S3.


\subsection*{Discussion}
Light harvesting complexes are fundamental components of the photosynthetic machinery on Earth. While there has been an extraordinary advance in the techniques to probe these systems at ultrafast time scales, it is conceptually unclear what truly quantum features with no classical counterpart these systems may exhibit and exploit for efficient energy transport \cite{Scholes12a}. Besides depending on long-range electronic interactions, excitation energy distribution is fundamentally affected by the molecular motions and harmonic (or in cases anharmonic) environments that modulate the electronic dynamics. We therefore put forward the idea that, precisely, investigation of non-classical phenomena associated to such molecular motions can pave the way towards understanding which non-trivial quantum phenomena can impact efficient energy distribution and trapping.  Within this line of thought, we investigate non-classical behaviour in a prototype dimer that conveys fundamental physical principles of vibration-assisted transport in a variety of light-harvesting antennae.  We demonstrate that quasi-resonances between excitonic transitions and underdamped high-energy intramolecular vibrations can trigger and harness non-trivial quantum behaviour of collective pigment motions that are initially in a thermal, fully incoherent, state.  In this scenario, correlations between non-classical fluctuations of collective pigment motions and efficient population of the target exciton state are found. Negative values both of the Mandel $Q-$parameter and of the quasi-probability distribution $P_w(\alpha)$ of the collective motion assure that no classical distribution can describe the same behaviour.  These non-trivial quantum phenomena are predicted for a variety of initial excitation conditions including statistical mixtures of excitons indicating that such non-classicality can be activated even under incoherent input of photo-excitations. Transient coherent ultrafast phonon spectroscopy \cite{Waldermann08} which is sensitive to low-phonon populations of high-energy vibrations \cite{Lee11} may provide an interesting experimental approach to probe the phenomena we describe. 

The prototype exciton-vibration dimers here investigated are representative of interband-like transfer pathways present in the majority of light-harvesting complexes. For instance in the LHCII complex, the considered dimer contributes to the fastest component of the $\textrm{Chl}_{b}\rightarrow \textrm{Chl}_{a}$ inter-band transfer pathway that directs excitation energy towards exit sites \cite{SchlauCohen09}. The demonstration that non-classicality is concomitant with efficient vibration-assisted transfer in these dimers therefore suggests that non-classical phenomena will have a contribution to the efficiency of the whole complex. A rigorous quantitative estimation of such contribution requires both a careful extension of our formalism to quantify these features in a multi-modal system (as likely other non-equilibrated vibrational motion will be involved) and a careful weighting of the vibration-assisted processes in the overall spatio-temporal distribution of energy.

The framework we propose can also be applied to gain insights into the non-classical response of vibrational motion in a variety of transport \cite{Novoderezhkin04a, Hay12} and sensing processes in biomolecules \cite{Dasgupta09, Turin96}. Of particular interest are charge transfer in reaction centres \cite{Novoderezhkin04a} and isomerization of photoreceptors \cite{Dasgupta09} where  specific intramolecular vibrational motions are known to be driven out of thermal equilibrium during the light-initiated electronic dynamics. It will also be interesting to use this framework to understand possible non-trivial quantum behaviour of molecular motions in chemical sensors \cite{Turin96} that are conjectured to operate through weak electronic interactions to sense molecular vibrations of the order of a thousand wavenumbers \cite{Brookes07, Franco11}.

We have also illustrated how in our prototype dimer with biologically relevant parameters, exciton-vibration dynamics can lead to non-exponential  excitonic energy distribution whereby dissipation into a low-energy thermal bath can be transiently prevented. From this view, coherent vibrational motions that do not relax quickly and whose fluctuations cannot be described classically may be seen as an internal quantum mechanism controlling energy distribution and storage. Further insights into the advantage of these non-trivial quantum behaviour may therefore be gained in a thermodynamic framework \cite{Jianming10,Dorfman13}. 

In conclusion, we have provided theoretical evidence that vibration-assisted exciton transport in prototype dimers, representative of interband-like transitions in a variety of photosythetic light-harvesting antennae, can exploit non-trivial quantum phenomena which cannot be reproduced by any classical counterpart, namely, non-classical fluctuations of collective pigment motions. Given that a variety of transport \cite{Novoderezhkin04a, Hay12}  and sensing phenomena \cite{Dasgupta09}  in biomolecules are known to involve non-equilibrium vibrational motion, our findings have broad implications for the field of quantum effects in biology as they suggest that investigating the non-classical nature of molecular fluctuations harnessed in these processes could be key to reveal a role for truly non-trivial quantum features.
\subsection*{Methods}
\footnotesize{
\textbf{Open quantum system dynamics.-}
To accurately account for the effects of the low-energy thermal bath we have adopted a hierarchical expansion of the exciton-vibration dynamics \cite{Tanimura89,Ishizaki09,Shi09}. We split the high-energy mode from the bath of harmonic oscillators and explicitly treat the quantum interactions between electronic excitations and these modes of frequency $\omega_\textrm{vib}$ by including it within the definition of the system, $H_{\textrm{ex}-\textrm{vib}}=H_\textrm{ex}\otimes \mathbbm{1}_\textrm{vib} + \mathbbm{1}_\textrm{ex}\otimes H_\textrm{vib} +H_\textrm{ex-vib}$. The electronic operators  then couple to the remaining vibrational modes $H_\textrm{I}=\sum_{i,\textbf{k}} g_i(\sigma_i^+\sigma_i^-\otimes\mathbbm{1}_\textrm{vib})(b_\textbf{k}^\dag+b_\textbf{k})$. This approach allows the effects of the low-energy thermal bath on the exciton-vibration dynamics to be accurately accounted for. Truncating the quantized mode at Fock level $M=6$ adequately describes both reduced dynamics of the collective quantized mode and the electronic dynamics of the prototype dimer at room temperature. A 
spectral density of the form $J(\omega) = 2\lambda\Omega_\textrm{c}\omega/(\Omega_c^2+\omega^2)$ is assumed,
where $\lambda$ and  $\Omega_\textrm{c}$ are the reorganisation energy and cut-off frequency respectively.
Supplementary Note 1 furnishes further details of the hierarchical expansion of exciton-vibration dynamics. Converged dynamics are obtained by terminating the hierarchical expansion at level $N=11$ and including just the $K=0$ Matsubara term. No additional Matsubara terms are necessary as $\Omega_c<K_\textrm{B}T$. For completeness we present dynamics including the $K=1$ term in Supplementary Fig. S4.

\textbf{Regularized quasi-probability distributions.}
The quasi-probability distribution we calculate is a regularized version of the $P$-representation
$
P_w(\alpha)=\frac{1}{\pi^2}\int \textrm{d}^2\xi~e^{\alpha\xi^*-\alpha^*\xi} \chi(\xi)\Omega_w(\xi)~,
$
where the quantum characteristic function $\chi(\xi)=\textrm{Tr}\{ e^{\xi b_{rd}^\dag-\xi^*b_{rd}}\varrho_\textrm{vib}\}$
 of the reduced state the vibration $\varrho_\textrm{vib}$ is reconstructed in the truncated Fock basis $\bra{m}\varrho_\textrm{vib}\ket{n}$. $\Omega_w(\xi)$ is a non-classicality filter \cite{Kiesel10} taken as the triangular function with $w=5$, which fulfils the necessary condition of a filter such that  $P_w(\xi)$ detects non-classicality. More details are given in Supplementary Note 2.}


\subsection*{Acknowledgements}
\noindent We thank  Greg Scholes and Rienk van Grondelle for discussions. We are also grateful with Carles Curuchet for a preliminary analysis on the specific molecular origins of vibrational modes in the PE545 complex and with Avinash Kolli for support on numerical calculations in the initial stages of this project. Financial support from the Engineering and Physical Sciences Research Council (EPSRC UK) Grant EP/G005222/1 and from the EU FP7 Project PAPETS (GA 323901)is gratefully acknowledged. 

\subsection*{Author contributions}
\noindent E. J. O  performed the calculations. A. O-C conceived, designed and supervised the research. Both authors analysed and discussed the results, and co-wrote the manuscript.

\subsection*{Competing financial interests} 
\noindent The authors declare no competing financial interests.

\end{document}